\pdfoutput=1

\documentclass[12pt,a4paper]{article}

\usepackage[colorinlistoftodos]{todonotes}

\usepackage{ifthen} 
\newboolean{pdflatex}
\setboolean{pdflatex}{true} 

\newboolean{articletitles}
\setboolean{articletitles}{true} 

\newboolean{uprightparticles}
\setboolean{uprightparticles}{false} 

\newboolean{inbibliography}
\setboolean{inbibliography}{false} 

\def\paperauthors{LHCb collaboration} 
\def\paperasciititle{Template for writing LHCb papers} 
\def\papertitle{Search for \CP violation in \LbTopK and \LbToppi decays} 
\def\paperkeywords{{High Energy Physics}, {LHCb}} 
\def\papercopyright{\the\year\ CERN for the benefit of the LHCb collaboration} 
\def\paperlicence{CC-BY-4.0 licence}
\def\paperlicenceurl{https://creativecommons.org/licenses/by/4.0/}

\usepackage[top=1in, bottom=1.25in, left=1in, right=1in]{geometry}
 
\usepackage{microtype}
\usepackage{lineno}  
\usepackage{xspace} 
\usepackage{caption} 

\usepackage{graphicx}  
\usepackage{color}
\usepackage{colortbl}
\graphicspath{{./figs/}} 
 
\usepackage{amsmath} 
\usepackage{amssymb}
\usepackage{amsfonts}
\usepackage{upgreek} 
 
\newcommand*\patchAmsMathEnvironmentForLineno[1]{%
\expandafter\let\csname old#1\expandafter\endcsname\csname #1\endcsname
\expandafter\let\csname oldend#1\expandafter\endcsname\csname
end#1\endcsname
 \renewenvironment{#1}%
   {\linenomath\csname old#1\endcsname}%
   {\csname oldend#1\endcsname\endlinenomath}%
}
\newcommand*\patchBothAmsMathEnvironmentsForLineno[1]{%
  \patchAmsMathEnvironmentForLineno{#1}%
  \patchAmsMathEnvironmentForLineno{#1*}%
}
\AtBeginDocument{%
\patchBothAmsMathEnvironmentsForLineno{equation}%
\patchBothAmsMathEnvironmentsForLineno{align}%
\patchBothAmsMathEnvironmentsForLineno{flalign}%
\patchBothAmsMathEnvironmentsForLineno{alignat}%
\patchBothAmsMathEnvironmentsForLineno{gather}%
\patchBothAmsMathEnvironmentsForLineno{multline}%
\patchBothAmsMathEnvironmentsForLineno{eqnarray}%
}
 
\usepackage{hyperxmp}
 
\usepackage[pdftex,
           pdfauthor={\paperauthors},
           pdftitle={\paperasciititle},
           pdfkeywords={\paperkeywords},
           pdfcopyright={Copyright (C) \papercopyright},
           pdflicenseurl={\paperlicenceurl}]{hyperref}

\usepackage[all]{hypcap} 
 
\usepackage{xspace} 
\usepackage{upgreek}

\def\lhcb   {\mbox{LHCb}\xspace}

\def\cdf    {\mbox{CDF}\xspace}

\def\MagUp {\mbox{\em Mag\kern -0.05em Up}\xspace}

\ifthenelse{\boolean{uprightparticles}}%
{

 \def\Ppi         {\ensuremath{\uppi}\xspace}

 \def\PDelta      {\ensuremath{\Delta}\xspace}                 
 \def\PXi      {\ensuremath{\Xi}\xspace}                 
 \def\PLambda      {\ensuremath{\Lambda}\xspace}                 
 \def\PSigma      {\ensuremath{\Sigma}\xspace}                 
 \def\POmega      {\ensuremath{\Omega}\xspace}                 
 \def\PUpsilon      {\ensuremath{\Upsilon}\xspace}                 
 
 \def\PB      {\ensuremath{\mathrm{B}}\xspace}                 
                  
 \def\PD      {\ensuremath{\mathrm{D}}\xspace}

 \def\PK      {\ensuremath{\mathrm{K}}\xspace}

 \def\Pb      {\ensuremath{\mathrm{b}}\xspace}                 
 \def\Pc      {\ensuremath{\mathrm{c}}\xspace}

 \def\Pi      {\ensuremath{\mathrm{i}}\xspace}

 \def\Pp      {\ensuremath{\mathrm{p}}\xspace}

 \def\Ps      {\ensuremath{\mathrm{s}}\xspace}

}
{

 \def\Ppi         {\ensuremath{\pi}\xspace}

 \mathchardef\PDelta="7101
 \mathchardef\PXi="7104
 \mathchardef\PLambda="7103
 \mathchardef\PSigma="7106
 \mathchardef\POmega="710A
 \mathchardef\PUpsilon="7107
                  
 \def\PB      {\ensuremath{B}\xspace}                 
                  
 \def\PD      {\ensuremath{D}\xspace}

 \def\PK      {\ensuremath{K}\xspace}

 \def\Pb      {\ensuremath{b}\xspace}                 
 \def\Pc      {\ensuremath{c}\xspace}

 \def\Pi      {\ensuremath{i}\xspace}

 \def\Pp      {\ensuremath{p}\xspace}

 \def\Ps      {\ensuremath{s}\xspace}

}

\makeatletter
\ifcase \@ptsize \relax
  \newcommand{\miniscule}{\@setfontsize\miniscule{4}{5}}
\or
  \newcommand{\miniscule}{\@setfontsize\miniscule{5}{6}}
\or
  \newcommand{\miniscule}{\@setfontsize\miniscule{5}{6}}
\fi
\makeatother

\DeclareRobustCommand{\optbar}[1]{\shortstack{{\miniscule (\rule[.5ex]{1.25em}{.18mm})}
  \\ [-.7ex] $#1$}}

\def\squark    {{\ensuremath{\Ps}}\xspace}

\def\cquark    {{\ensuremath{\Pc}}\xspace}

\def\bquark    {{\ensuremath{\Pb}}\xspace}

\def\pion   {{\ensuremath{\Ppi}}\xspace}

\def\pip    {{\ensuremath{\pion^+}}\xspace}
\def\pim    {{\ensuremath{\pion^-}}\xspace}

\def\kaon    {{\ensuremath{\PK}}\xspace}
  \def\Kbar    {{\kern 0.2em\overline{\kern -0.2em \PK}{}}\xspace}

\def\KorKbar    {\kern 0.18em\optbar{\kern -0.18em K}{}\xspace}
\def\Kz      {{\ensuremath{\kaon^0}}\xspace}

\def\Kp      {{\ensuremath{\kaon^+}}\xspace}
\def\Km      {{\ensuremath{\kaon^-}}\xspace}

\def\KS      {{\ensuremath{\kaon^0_{\mathrm{ \scriptscriptstyle S}}}}\xspace}

  \def\Dbar    {{\kern 0.2em\overline{\kern -0.2em \PD}{}}\xspace}
\def\D       {{\ensuremath{\PD}}\xspace}

\def\DorDbar    {\kern 0.18em\optbar{\kern -0.18em D}{}\xspace}
\def\Dz      {{\ensuremath{\D^0}}\xspace}

\def\Dp      {{\ensuremath{\D^+}}\xspace}

\def\Dstarp  {{\ensuremath{\D^{*+}}}\xspace}

\def\B       {{\ensuremath{\PB}}\xspace}
\def\Bbar    {{\ensuremath{\kern 0.18em\overline{\kern -0.18em \PB}{}}}\xspace}

\def\BorBbar    {\kern 0.18em\optbar{\kern -0.18em B}{}\xspace}
\def\Bz      {{\ensuremath{\B^0}}\xspace}

\def\Bu      {{\ensuremath{\B^+}}\xspace}

\def\Bp      {{\ensuremath{\Bu}}\xspace}

\def\Bd      {{\ensuremath{\B^0}}\xspace}
\def\Bs      {{\ensuremath{\B^0_\squark}}\xspace}

  \def\Y#1S{\ensuremath{\PUpsilon{(#1S)}}\xspace}

\def\proton      {{\ensuremath{\Pp}}\xspace}
\def\antiproton  {{\ensuremath{\overline \proton}}\xspace}

\def\Lz          {{\ensuremath{\PLambda}}\xspace}
\def\Lbar        {{\ensuremath{\kern 0.1em\overline{\kern -0.1em\PLambda}}}\xspace}
\def\LorLbar    {\kern 0.18em\optbar{\kern -0.18em \PLambda}{}\xspace}

\def\Lb      {{\ensuremath{\Lz^0_\bquark}}\xspace}
\def\Lbbar   {{\ensuremath{\Lbar{}^0_\bquark}}\xspace}
\def\Lc      {{\ensuremath{\Lz^+_\cquark}}\xspace}

\newcommand{\decay}[2]{\ensuremath{#1\!\to #2}\xspace}         

\def\to                 {\ensuremath{\rightarrow}\xspace}

\def\CP                {{\ensuremath{C\!P}}\xspace}

\def\BdToKpi      {\decay{\Bd}{\Kp\pim}}

\def\LbTopK       {\decay{\Lb}{\proton\Km}}

\def\LbToppi      {\decay{\Lb}{\proton\pim}}

\def\AT#1     {\ensuremath{A_{\mathrm{T}}^{#1}}\xspace}           

\def\C#1      {\ensuremath{\mathcal{C}_{#1}}\xspace}                       
\def\Cp#1     {\ensuremath{\mathcal{C}_{#1}^{'}}\xspace}                    
\def\Ceff#1   {\ensuremath{\mathcal{C}_{#1}^{\mathrm{(eff)}}}\xspace}        
\def\Cpeff#1  {\ensuremath{\mathcal{C}_{#1}^{'\mathrm{(eff)}}}\xspace}       
\def\Ope#1    {\ensuremath{\mathcal{O}_{#1}}\xspace}                       
\def\Opep#1   {\ensuremath{\mathcal{O}_{#1}^{'}}\xspace}                    

\newcommand{\tev}{\ifthenelse{\boolean{inbibliography}}{\ensuremath{~T\kern -0.05em eV}}{\ensuremath{\mathrm{\,Te\kern -0.1em V}}}\xspace}
\newcommand{\gev}{\ensuremath{\mathrm{\,Ge\kern -0.1em V}}\xspace}
\newcommand{\mev}{\ensuremath{\mathrm{\,Me\kern -0.1em V}}\xspace}
\newcommand{\kev}{\ensuremath{\mathrm{\,ke\kern -0.1em V}}\xspace}
\newcommand{\ev}{\ensuremath{\mathrm{\,e\kern -0.1em V}}\xspace}
\newcommand{\gevc}{\ensuremath{{\mathrm{\,Ge\kern -0.1em V\!/}c}}\xspace}
\newcommand{\mevc}{\ensuremath{{\mathrm{\,Me\kern -0.1em V\!/}c}}\xspace}
\newcommand{\gevcc}{\ensuremath{{\mathrm{\,Ge\kern -0.1em V\!/}c^2}}\xspace}
\newcommand{\gevgevcccc}{\ensuremath{{\mathrm{\,Ge\kern -0.1em V^2\!/}c^4}}\xspace}
\newcommand{\mevcc}{\ensuremath{{\mathrm{\,Me\kern -0.1em V\!/}c^2}}\xspace}

\def\mum  {\ensuremath{{\,\upmu\mathrm{m}}}\xspace}

\def\invfb   {\ensuremath{\mbox{\,fb}^{-1}}\xspace}

\newcommand{\chisqip}{\ensuremath{\chi^2_{\text{IP}}}\xspace}

\def\gsim{{~\raise.15em\hbox{$>$}\kern-.85em
          \lower.35em\hbox{$\sim$}~}\xspace}
\def\lsim{{~\raise.15em\hbox{$<$}\kern-.85em
          \lower.35em\hbox{$\sim$}~}\xspace}

\def\pt         {\ensuremath{p_{\mathrm{ T}}}\xspace}
\def\ptot       {\ensuremath{p}\xspace}

\def\evtgen     {\mbox{\textsc{EvtGen}}\xspace}

\def\geant      {\mbox{\textsc{Geant4}}\xspace}

\def\photos     {\mbox{\textsc{Photos}}\xspace}

\def\pythia     {\mbox{\textsc{Pythia}}\xspace}

\def\tell1  {TELL1\xspace}
\def\ukl1   {UKL1\xspace}

\def\AD {\ensuremath{A_{\mathrm{D}}}\xspace}
\def\APLb {\ensuremath{A_{\mathrm P}^\Lb}\xspace}

\def\ADKz {\ensuremath{A_{\mathrm D}^{\Kz}}\xspace}

\def\ADKm {\ensuremath{A_{\mathrm D}^{\Km}}\xspace}

\def\ADpim {\ensuremath{A_{\mathrm D}^{\pim}}\xspace}
\def\ADp {\ensuremath{A_{\mathrm D}^\proton}\xspace}

\def\Araw {\ensuremath{A_{\mathrm{raw}}}\xspace}
\def\dacp {\ensuremath{\Delta A_{\CP}}\xspace}
\def\ACPpK {\ensuremath{A_\CP^{\proton \Km}}\xspace}
\def\ACPppi {\ensuremath{A_\CP^{\proton \pim}}\xspace}
\def\ArawpK {\ensuremath{A_\mathrm{raw}^{\proton \Km}}\xspace}
\def\Arawppi {\ensuremath{A_\mathrm{raw}^{\proton \pim}}\xspace}
\def\APID {\ensuremath{A_\mathrm{PID}}\xspace}
\def\APIDpK {\ensuremath{A_\mathrm{PID}^{\proton \Km}}\xspace}
\def\APIDppi {\ensuremath{A_\mathrm{PID}^{\proton \pim}}\xspace}
\def\ATpK  {\ensuremath{A_\mathrm{trigger}^{\proton \Km}}\xspace}
\def\ATppi  {\ensuremath{A_\mathrm{trigger}^{\proton \pim}}\xspace}
\def\SpK  {\ensuremath{S_{\proton \Km}}\xspace}
\def\Sppi  {\ensuremath{S_{\proton \pim}}\xspace}
 
\usepackage{cite} 
\usepackage{mciteplus}
\usepackage{multirow}
\usepackage{csquotes}

\usepackage{tabularx}
\newcolumntype{L}[1]{>{\raggedright\arraybackslash}p{#1}}
\newcolumntype{C}[1]{>{\centering\arraybackslash}p{#1}}
\newcolumntype{R}[1]{>{\raggedleft\arraybackslash}p{#1}}

\RequirePackage[normalem]{ulem} 
\RequirePackage{color}\definecolor{RED}{rgb}{1,0,0}\definecolor{BLUE}{rgb}{0,0,1} 

\usepackage{longtable} 

\begin{document}

\renewcommand{\thefootnote}{\fnsymbol{footnote}}
\setcounter{footnote}{1}


\begin{titlepage}
\pagenumbering{roman}

\vspace*{-1.5cm}
\centerline{\large EUROPEAN ORGANIZATION FOR NUCLEAR RESEARCH (CERN)}
\vspace*{1.5cm}
\noindent
\begin{tabular*}{\linewidth}{lc@{\extracolsep{\fill}}r@{\extracolsep{0pt}}}
\ifthenelse{\boolean{pdflatex}}
{\vspace*{-1.5cm}\mbox{\!\!\!\includegraphics[width=.14\textwidth]{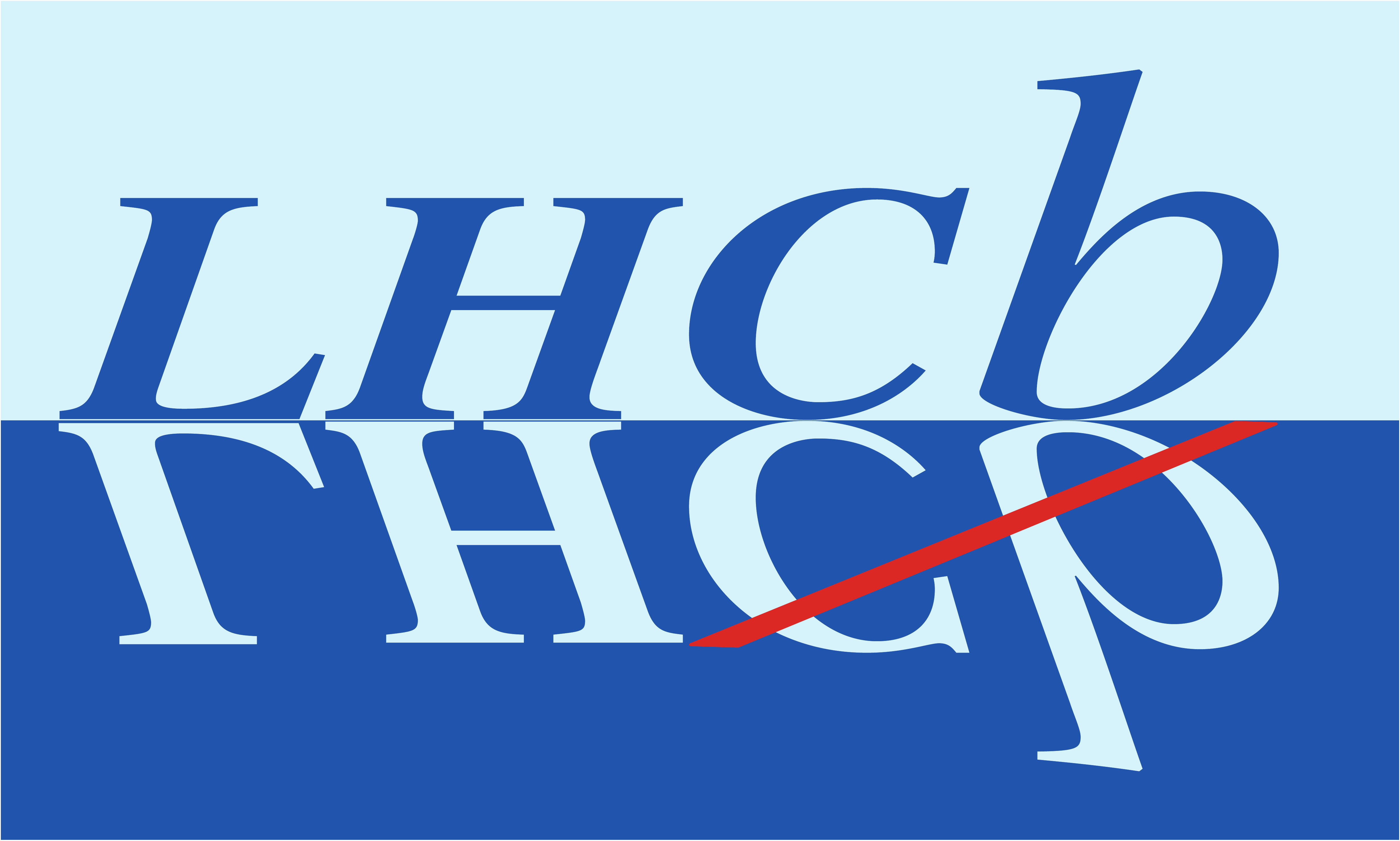}} & &}%
{\vspace*{-1.2cm}\mbox{\!\!\!\includegraphics[width=.12\textwidth]{lhcb-logo.eps}} & &}%
\\
 & & CERN-EP-2018-189 \\  
 & & LHCb-PAPER-2018-025 \\  
 & & July 17, 2018 \\ 
 & & \\
\end{tabular*}

\vspace*{4.0cm}

{\normalfont\bfseries\boldmath\huge
\begin{center}
  \papertitle
\end{center}
}

\vspace*{2.0cm}

\begin{center}
\paperauthors\footnote{Authors are listed at the end of this paper.}
\end{center}

\vspace{\fill}

\begin{abstract}
  \noindent
A search for \CP violation in \LbTopK and \LbToppi decays is presented using a sample of $pp$ collisions collected with the LHCb detector and corresponding to an integrated luminosity of 3.0\invfb. The \CP-violating asymmetries are measured to be $\ACPpK = -0.020 \pm 0.013\pm  0.019$ and $\ACPppi = -0.035 \pm 0.017 \pm 0.020 $, and their difference $\ACPpK-\ACPppi = 0.014 \pm 0.022 \pm 0.010$, where the first uncertainties are statistical and the second systematic. These are the most precise measurements of such asymmetries to date.

\end{abstract}

\vspace*{2.0cm}

\begin{center}
   Published in Phys. Lett. B 787 (2018) 124-133
\end{center}

\vspace{\fill}

{\footnotesize
\centerline{\copyright~\papercopyright. \href{\paperlicenceurl}{\paperlicence}.}}
\vspace*{2mm}

\end{titlepage}


\newpage
\setcounter{page}{2}
\mbox{~}
%
%
%
%

\cleardoublepage


\renewcommand{\thefootnote}{\arabic{footnote}}
\setcounter{footnote}{0}



\pagestyle{plain} 
\setcounter{page}{1}
\pagenumbering{arabic}


%

\section{Introduction}
\label{sec:Introduction}
\noindent
The non-invariance of weak interactions under the combined application of charge conjugation ($C$) and parity ($P$) transformations is accommodated within the Standard Model by the Cabibbo-Kobayashi-Maskawa mechanism~\cite{Cabibbo:1963yz,Kobayashi:1973fv}. The violation of the \CP symmetry was discovered in neutral-kaon decays~\cite{Christenson:1964fg}, and later observed with \Bz~\cite{Aubert:2001nu, Abe:2001xe,Aubert:2004qm,Abe:2004us,Aubert:2007mj,Lees:2012mma,Adachi:2013mae,Duh:2012ie,Aaltonen:2014vra}, \Bp~\cite{LHCb-PAPER-2012-001} and \Bs mesons~\cite{LHCb-PAPER-2013-018,Aaltonen:2014vra}. First evidence for \CP violation in the \bquark-baryon sector was found more recently~\cite{LHCb-PAPER-2016-030}. The decays \LbTopK and \LbToppi are mediated by the same quark-level transitions contributing to charmless two-body \Bz and \Bs decays to charged pions and kaons, where nonzero values of the \CP asymmetries are well established~\cite{LHCb-PAPER-2013-018}. The inclusion of charge-conjugate processes is implied throughout.

Predictions for the \CP asymmetries in the decays of the \Lb baryon to two-body charmless final states $\proton \Km$ or $\proton \pim$ range from a few percent in the generalised factorisation approach~\cite{Hsiao:2014mua,Ali:1998eb} up to approximately $30\%$
 within the perturbative quantum-chromodynamics formalism~\cite{Lu:2009cm}. The only measurements of these quantities available to date were performed by the \cdf collaboration~\cite{Aaltonen:2014vra}. The asymmetries were found to be compatible with zero within an uncertainty of $8$ to $9\%$.

This Letter reports on a search for \CP violation in \LbTopK and \LbToppi decays, using $pp$-collision data collected with the LHCb detector at centre-of-mass energies of 7 and 8\tev and corresponding to 3.0\invfb of integrated luminosity.
The \CP asymmetry is defined as
\begin{equation}
A_\CP^f \equiv \frac{\Gamma(\Lb \to f) - \Gamma(\Lbbar \to \overline{f})}{\Gamma(\Lb \to f) + \Gamma(\Lbbar \to \overline{f})},
\end{equation}
where $\Gamma$ is the partial width of the given decay, with $f \equiv \proton\Km$~$(\proton \pim)$ and $\overline{f} \equiv \antiproton\Kp$~$(\antiproton \pip)$. In addition, the difference of the two \CP asymmetries, $\dacp \equiv \ACPpK - \ACPppi$, is also reported. As the main systematic uncertainties cancel in the difference, this quantity will become useful with the increasing size of the data sample.

The Letter is organised as follows. After a brief introduction on the detector, trigger and simulation in Sec.~\ref{sec:DetectorTriggerAndSimulation}, the formalism needed to relate the physical \CP asymmetries to the experimental measurements is presented in Sec.~\ref{sec:Formalism}. Then, the event selection and the invariant-mass fit are described in Secs.~\ref{sec:EventSelection} and~\ref{sec:InvariantMassFit}, respectively. The determination of instrumental asymmetries and systematic uncertainties is discussed in Sec.~\ref{sec:InstrumentalAsymmetriesAndSystematicUncertainties}. Finally, results are given and conclusions are drawn in Sec.~\ref{sec:ResultsAndConclusions}.

\section{Detector, trigger and simulation}
\label{sec:DetectorTriggerAndSimulation}
The \lhcb detector~\cite{Alves:2008zz,LHCb-DP-2014-002} is a single-arm forward
spectrometer covering the \mbox{pseudorapidity} range $2<\eta <5$,
designed for the study of particles containing \bquark or \cquark
quarks. The detector includes a high-precision tracking system
consisting of a silicon-strip vertex detector surrounding the $pp$
interaction region~\cite{LHCb-DP-2014-001}, a large-area silicon-strip detector located
upstream of a dipole magnet with a bending power of about
$4{\mathrm{\,Tm}}$, and three stations of silicon-strip detectors and straw
drift tubes~\cite{LHCb-DP-2013-003} placed downstream of the magnet.
The tracking system provides a measurement of the momentum, \ptot, of charged particles with
a relative uncertainty that varies from 0.5\% at low momentum to 1.0\% at 200\gevc.
The minimum distance of a track to a primary vertex (PV), the impact parameter (IP),
is measured with a resolution of $(15+29/\pt)\mum$,
where \pt is the component of the momentum transverse to the beam, in\,\gevc.
Different types of charged hadrons are distinguished using information
from two ring-imaging Cherenkov detectors~\cite{LHCb-DP-2012-003}.
Photons, electrons and hadrons are identified by a calorimeter system consisting of
scintillating-pad and preshower detectors, an electromagnetic
calorimeter and a hadronic calorimeter. Muons are identified by a
system composed of alternating layers of iron and multiwire
proportional chambers~\cite{LHCb-DP-2012-002}.
The online event selection is performed by a trigger~\cite{LHCb-DP-2012-004},
which consists of a hardware stage, based on information from the calorimeter and muon
systems, followed by a software stage, which applies a full event
reconstruction.

 Simulated events are used to study the modelling of various mass line shapes. In the simulation, proton-proton collisions are generated using \pythia~\cite{Sjostrand:2007gs} with a specific \lhcb configuration~\cite{LHCb-PROC-2010-056}. Decays of hadronic particles are described by \evtgen~\cite{Lange:2001uf}, in which final-state radiation is generated using \photos~\cite{Golonka:2005pn}. The interaction of the generated particles with the detector and its response are implemented using the \geant toolkit~\cite{Allison:2006ve, *Agostinelli:2002hh} as described in Ref.~\cite{LHCb-PROC-2011-006}.

\section{Formalism}
\label{sec:Formalism}
The \CP asymmetries of \LbTopK and \LbToppi decays are approximated as the sums of various experimental quantities
\begin{eqnarray}
\ACPpK  &=& \ArawpK - \ADp - \ADKm - \APIDpK - \APLb - \ATpK, \label{eq:ACPpK}\\
\ACPppi &=& \Arawppi - \ADp - \ADpim - \APIDppi - \APLb - \ATppi, \label{eq:ACPppi}
\end{eqnarray}
where $\Araw^f$ is the measured raw asymmetry between the yields of the decays $\Lb \to f$ and $\Lbbar \to \overline{f}$, with $f=\proton \Km$~$(\proton \pim)$ and $\overline{f}$ its charge conjugate; $\AD^h$ is the asymmetry between the detection efficiencies for particle $h$ and its charge conjugate, with $h=p,\ \Km$ or $\pim$; the symbol $\APID^f$ stands for the asymmetry between the particle-identification (PID) efficiencies for the final states $f$ and $\overline{f}$; \APLb is the asymmetry between the production cross-sections of \Lb and \Lbbar baryons; and $A^f_\mathrm{trigger}$ is the asymmetry between the trigger efficiencies for the particles in the final states $f$ and $\overline{f}$. This linear approximation is valid to a good enough accuracy due to the smallness of the terms involved.

The raw asymmetry is defined as
\begin{equation}
\Araw^f \equiv \frac{N(\Lb \to f) - N(\Lbbar \to \overline{f})}{N(\Lb \to f) + N(\Lbbar \to \overline{f})},
\end{equation}
where $N$ denotes the observed signal yield for the given decay, obtained in this analysis by means of extended binned maximum-likelihood fits to the  $\proton \Km$ and $\proton \pim$ invariant-mass spectra.

The proton, kaon and pion detection asymmetries are defined as
\begin{equation}
\label{eq:DetectionAsymmetries}
\ADp \equiv \frac{\varepsilon_\mathrm{rec}^{p}-\varepsilon_\mathrm{rec}^{\antiproton}}{\varepsilon_\mathrm{rec}^{p}+\varepsilon_\mathrm{rec}^{\antiproton}},\ \ \ \ \ADKm \equiv \frac{\varepsilon_\mathrm{rec}^{\Km}-\varepsilon_\mathrm{rec}^{\Kp}}{\varepsilon_\mathrm{rec}^{\Km}+\varepsilon_\mathrm{rec}^{\Kp}}, \ \ \ \ \   \ADpim \equiv \frac{\varepsilon_\mathrm{rec}^{\pim}-\varepsilon_\mathrm{rec}^{\pip}}{\varepsilon_\mathrm{rec}^{\pim}+\varepsilon_\mathrm{rec}^{\pip}},
\end{equation}
where $\varepsilon_\mathrm{rec}$ is the total efficiency to reconstruct the given particle, excluding PID. Such asymmetries are mostly due to the different interaction cross-sections of particles and antiparticles with the detector material.
The kaon and pion detection asymmetries are measured using charm-meson control samples employing the procedures described in Refs.~\cite{LHCb-PAPER-2014-013,LHCb-PAPER-2012-009}.
The kaon detection asymmetry is obtained by subtracting the raw asymmetries of the $\Dp \to \KS\pip$ and $\Dp \to \Km\pip\pip$ decay modes and correcting for the $\Kz$ (\ADKz)~\cite{LHCb-PAPER-2014-013} and pion detection asymmetries. The latter is measured from the ratio of partially to fully reconstructed $\Dstarp \to \Dz(\to \Km\pip\pip\pim)\pip$ decays. The proton detection asymmetry is obtained from simulated events.

The PID asymmetries are measured from large calibration samples and are defined as
\begin{equation}
\label{eq:PIDAsym}
\APID^f \equiv \frac{\varepsilon_\mathrm{PID}^f-\varepsilon_\mathrm{PID}^{\overline{f}}}{\varepsilon_\mathrm{PID}^f+\varepsilon_\mathrm{PID}^{\overline{f}}},
\end{equation}
where $\varepsilon_\mathrm{PID}^{f\,(\overline{f})}$ is the PID efficiency for a final state $f\,(\overline{f})$ given a set of PID requirements.

The \Lb production asymmetry is defined as
\begin{equation}
\APLb = \frac{\sigma(\Lb) - \sigma(\Lbbar)}{\sigma(\Lb) + \sigma(\Lbbar)},
\end{equation}
where $\sigma$ denotes the inclusive production cross-section in the LHCb acceptance. The production asymmetry is taken as an external input, following Ref.~\cite{LHCb-PAPER-2016-062}.

Finally, asymmetries may arise if the hardware and software trigger used to collect data do not have the same efficiencies on oppositely charged particles. These effects are estimated through various data-driven techniques, as described in Sec.~\ref{sec:InstrumentalAsymmetriesAndSystematicUncertainties}.

\section{Event selection}
\label{sec:EventSelection}
The event selection starts with the reconstruction of \bquark hadrons formed by two oppositely charged tracks with $\pt > 1\gevc$, inconsistent with originating from any PV and required to form a common vertex. Each \bquark-hadron candidate needs to have a transverse momentum greater than $1.2\gevc$ and an invariant mass, computed assigning the pion mass to both daughter tracks, in the range between 4.8 and 5.8\gevcc. Finally, each \bquark-hadron candidate is required to be consistent with originating from a PV.

Particle-identification selection criteria are applied to divide the data sample into mutually exclusive subsamples corresponding to the final-state hypotheses $\proton \Km$, $\antiproton \Kp$, $\proton \pim$, $\antiproton \pip$, $\Kp\pim$, $\Km\pip$, $\Kp\Km$ and $\pip\pim$. The latter four combinations are selected to study the background due to two-body \B decays, where one or both final-state particles are misidentified.

The event selection is further refined using a boosted decision tree (BDT) classifier~\cite{Roe,Breiman} to reject combinatorial background. This algorithm combines the information from several input quantities to obtain a discriminant variable used to classify the \bquark-hadron candidates as signal or background. The following properties of the final-state particles are used as input variables: the transverse momentum of the \bquark-hadron decay products, the logarithms of their \chisqip values, where \chisqip is defined as the difference in the vertex-fit $\chi^2$ of a given PV reconstructed with and without the candidate under consideration, the quality of the common vertex fit of the two tracks and the distance of closest approach between the two tracks. The BDT also exploits the following properties of the \bquark-hadron candidate: the transverse momentum, the \chisqip quantity, and the logarithm of the flight distance with respect to the associated PV, defined as that with the smallest \chisqip with respect to the \bquark-hadron candidate. The BDT is trained using simulated signal decays and combinatorial background events from data in the high-mass sideband.

The selection criteria on the BDT classifier and the PID variables are optimised separately for the \LbTopK and \LbToppi decays. Two different selections, denoted hereafter as \SpK and \Sppi, are aimed at obtaining the best statistical sensitivity on each of the two \CP asymmetries. Common PID requirements are used for the final states containing only kaons and pions. Multiple candidates are present in less than 0.05\% of the events in the final sample. Only one candidate is accepted for each event on the basis of a reproducible pseudorandom sequence.

\section{Invariant-mass fit}
\label{sec:InvariantMassFit}
For each final-state hypothesis, namely $\proton \Km$, $\antiproton \Kp$, $\proton \pim$, $\antiproton \pip$, $\Kp\pim$, $\Km\pip$, $\Kp\Km$ and $\pip\pim$, the invariant-mass distribution of selected candidates is modelled by an appropriate probability density function. These models are used to perform a simultaneous fit to the eight invariant-mass spectra and determine at once the yields of all two-body \mbox{\bquark-hadron} decays contributing to the spectra.
Three categories are considered for the background: combinatorial, due to random association of tracks; partially reconstructed, due to multibody \bquark-hadron decays with one or more particles not reconstructed; and cross-feed, arising from other two-body \bquark-hadron decays where one or both final-state particles are misidentified.

The model used to describe each signal is obtained by convolving the sum of two Gaussian functions with common mean, accounting for mass-resolution effects, with a power-law function that accounts for final-state photon radiation effects. The power-law distribution is taken from analytical quantum-electrodynamics calculations~\cite{Baracchini:2005wp} and the correctness of the model is checked against simulated events generated with \photos~\cite{Golonka:2005pn}. The parameter governing the tail due to final-state photon radiation effects is different for each decay mode. This model describes well the invariant-mass distributions predicted by the simulation.

The combinatorial background is modelled using exponential functions. The partially reconstructed background is parameterised using ARGUS functions~\cite{Albrecht:1994aa} convolved with the sum of two Gaussian functions with zero mean values, whose relative fraction and widths are in common with the signal model. Finally, the cross-feed background is modelled using simulated two-body \bquark-hadron decays and a kernel estimation method~\cite{Cranmer:2001aa}.
The cross-feed background yields are set to the corresponding two-body \bquark-hadron decay yields, determined by the simultaneous fit, multiplied by appropriate PID efficiency ratios. The efficiencies for a given PID requirement are obtained from large calibration samples of $\Dstarp \to \Dz(\to \Km\pip)\pip$, $\Lz \to \proton \pim$ and $\Lc \to \proton \Km \pip$ decays, with the aid of simulated events in the case of protons to account for phase-space regions not covered by the calibration samples (about 20\% of the protons from signal decays). The efficiencies are determined in bins of particle momentum, pseudorapidity and track multiplicity, as the performances of the RICH detectors depend on such variables. They are then averaged over the momentum and pseudorapidity distributions of the final-state particles and over the distribution of track multiplicity in selected events.

\begin{figure}[!tbp]
\includegraphics[width=0.49\textwidth]{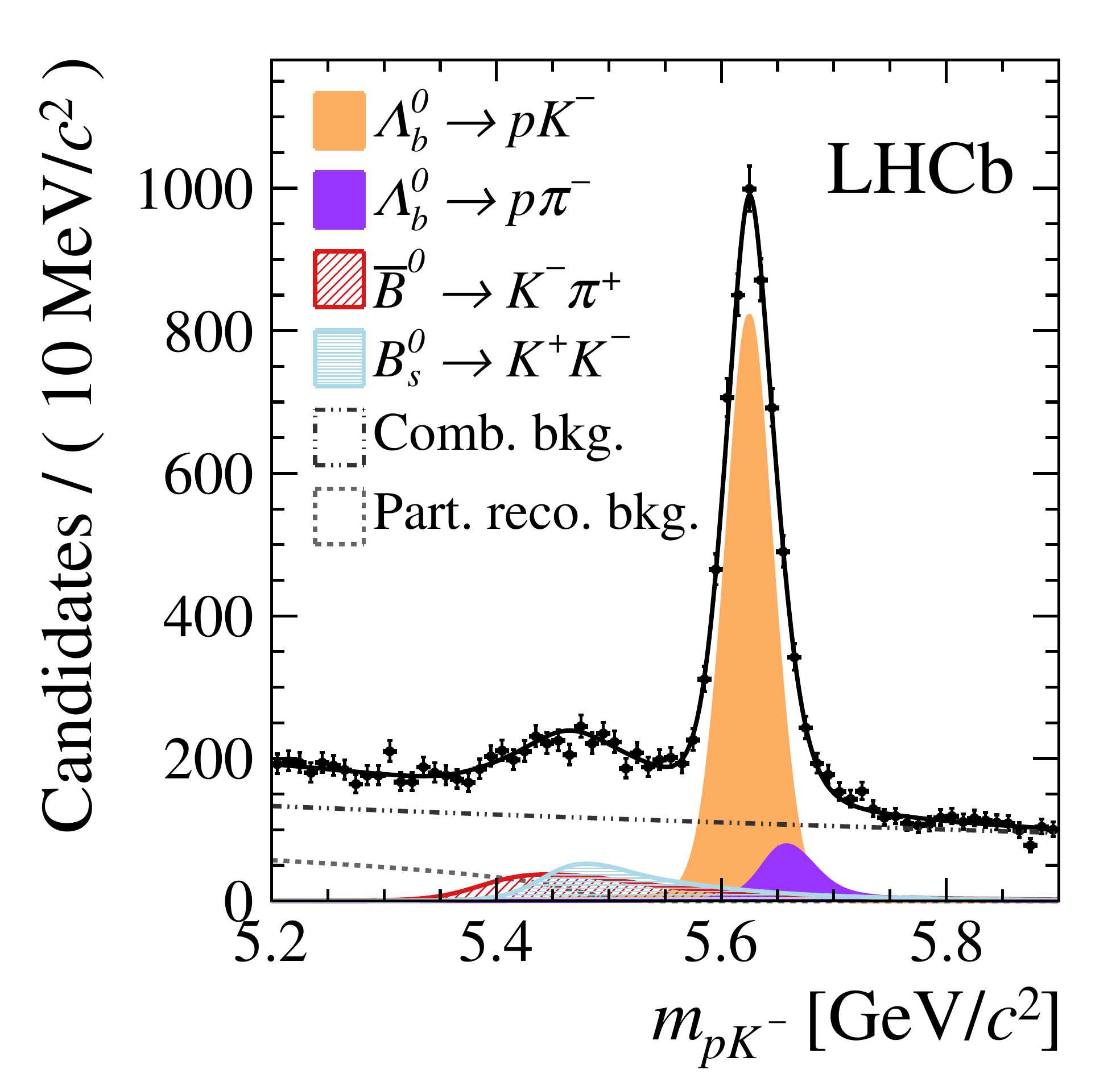}
\includegraphics[width=0.49\textwidth]{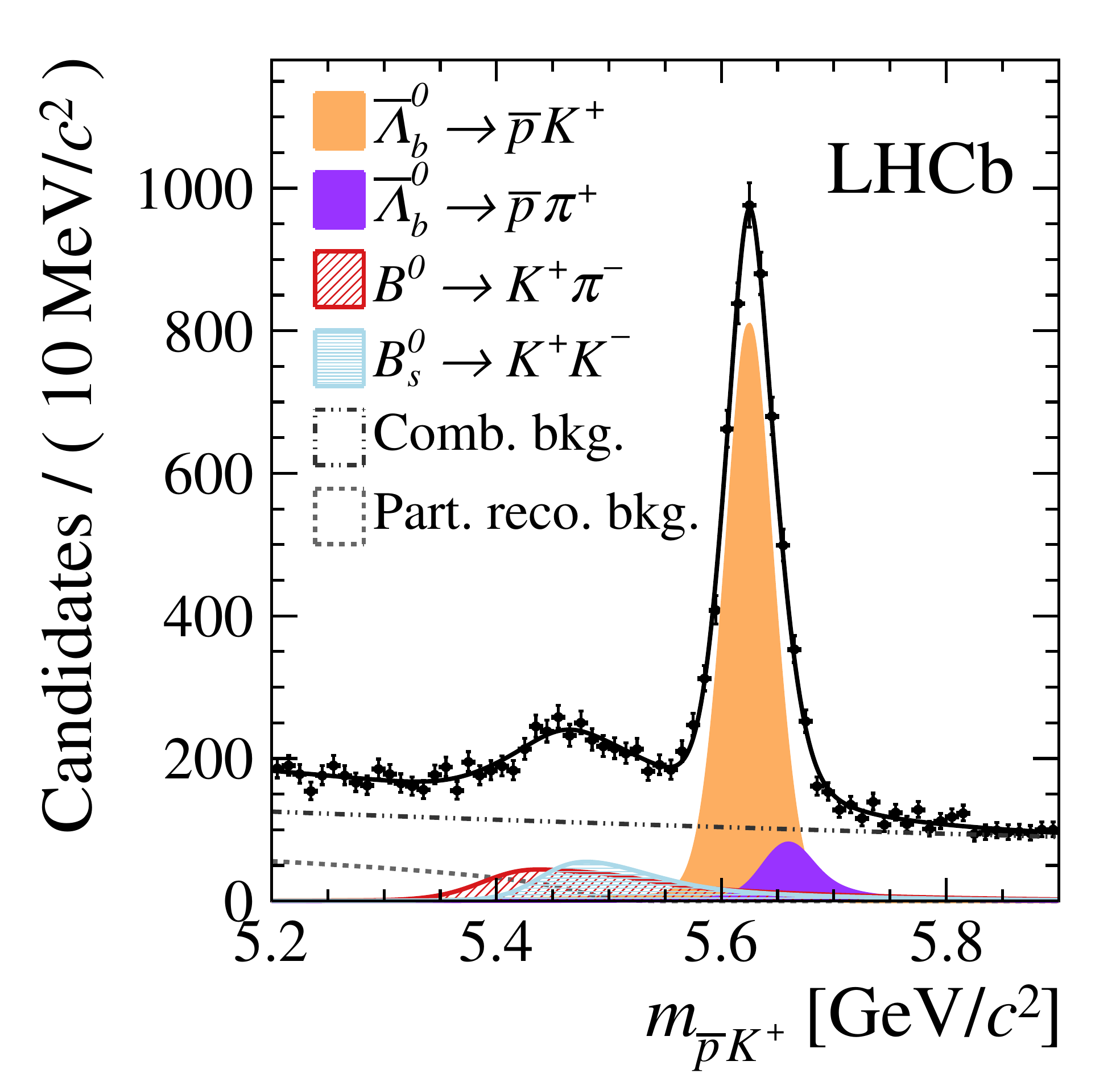}\\
\includegraphics[width=0.49\textwidth]{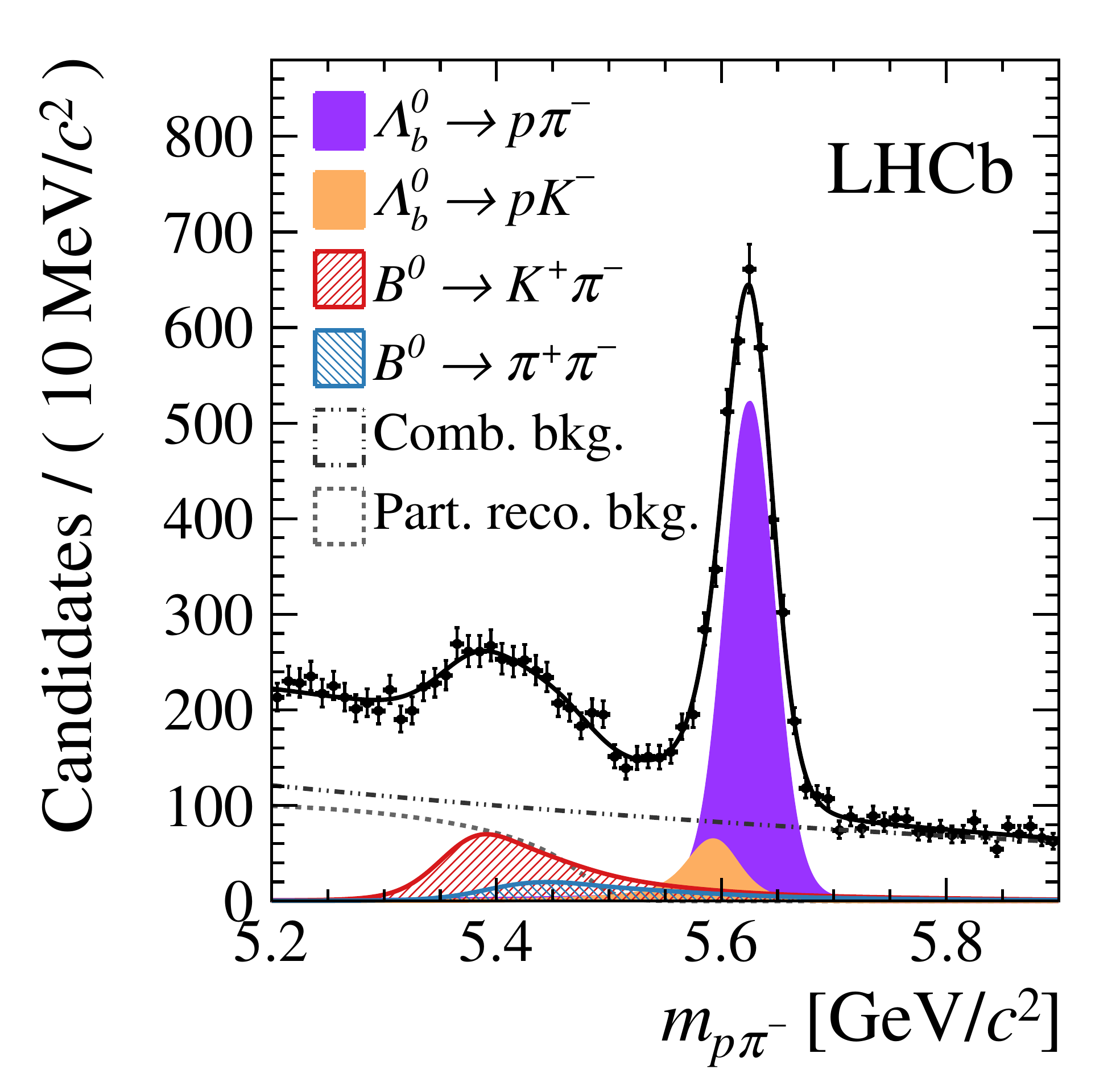}
\includegraphics[width=0.49\textwidth]{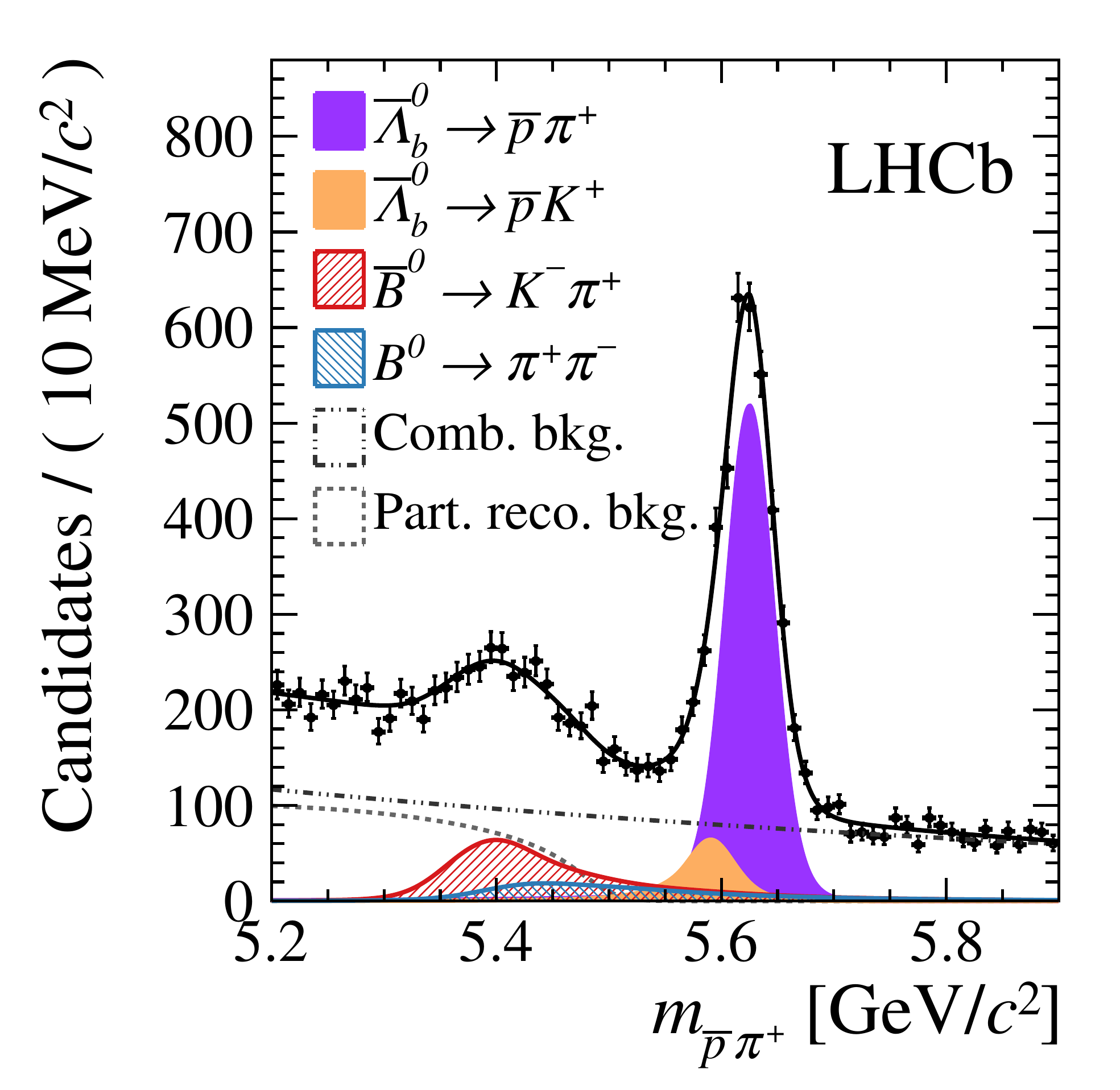}
\caption{Invariant-mass distributions: (top left) $m_{\proton \Km}$, (top right) $m_{\antiproton \Kp}$, (bottom left) $m_{\proton \pim}$ and (bottom right) $m_{\antiproton \pip}$ for candidates passing the (top) \SpK and (bottom) \Sppi selections. The results of the fits are superimposed.}
\label{fig:FinalFits}
\end{figure}

After the application of the optimal BDT and PID requirements, an extended binned maximum-likelihood fit with a bin width of 5\mevcc is performed simultaneously to the eight two-body invariant-mass spectra for each of the two selections, \SpK and \Sppi. The $m_{\proton \Km}$ and $m_{\proton \pim}$ invariant-mass distributions are shown in Fig.~\ref{fig:FinalFits}, with the results of the fits superimposed.
The values of the raw asymmetries and of the signal yields obtained from the fits to the candidates passing the respective \SpK or \Sppi selection are $\ArawpK=(1.0\pm1.3)\%$, $\Arawppi=(0.5\pm1.7)\%$, $N^{\proton\Km}_\mathrm{sig}+N^{\antiproton\Kp}_\mathrm{sig} = 8\,847\pm125$ and $N^{\proton\pim}_\mathrm{sig}+N^{\antiproton\pip}_\mathrm{sig} = 6\,026 \pm 105$.

The fit is validated by generating a large number of pseudoexperimental data samples according to the total probability density function of the model and performing an extended binned maximum-likelihood fit to each sample. The resulting pull distributions for \ArawpK and \Arawppi are found to be Gaussian with zero means and unitary widths.

\section{Instrumental asymmetries and systematic uncertainties}
\label{sec:InstrumentalAsymmetriesAndSystematicUncertainties}
The determination of the instrumental asymmetries introduced in Eqs.~\eqref{eq:ACPpK} and~\eqref{eq:ACPppi} is crucial to obtain the \CP asymmetries, as described in Sec.~\ref{sec:Formalism}.

\begin{figure}[!tbp]
\includegraphics[width=0.49\textwidth]{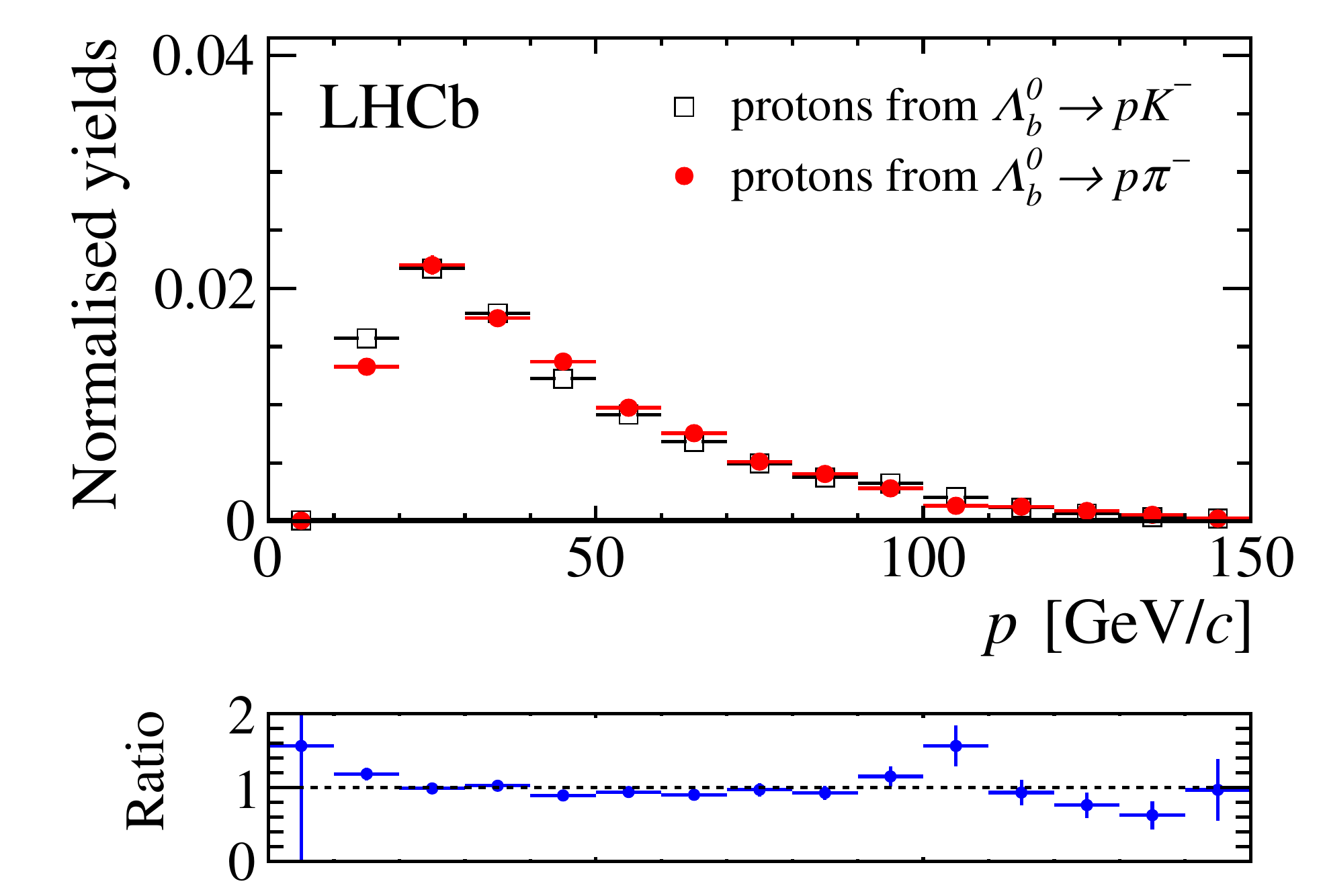}
\includegraphics[width=0.49\textwidth]{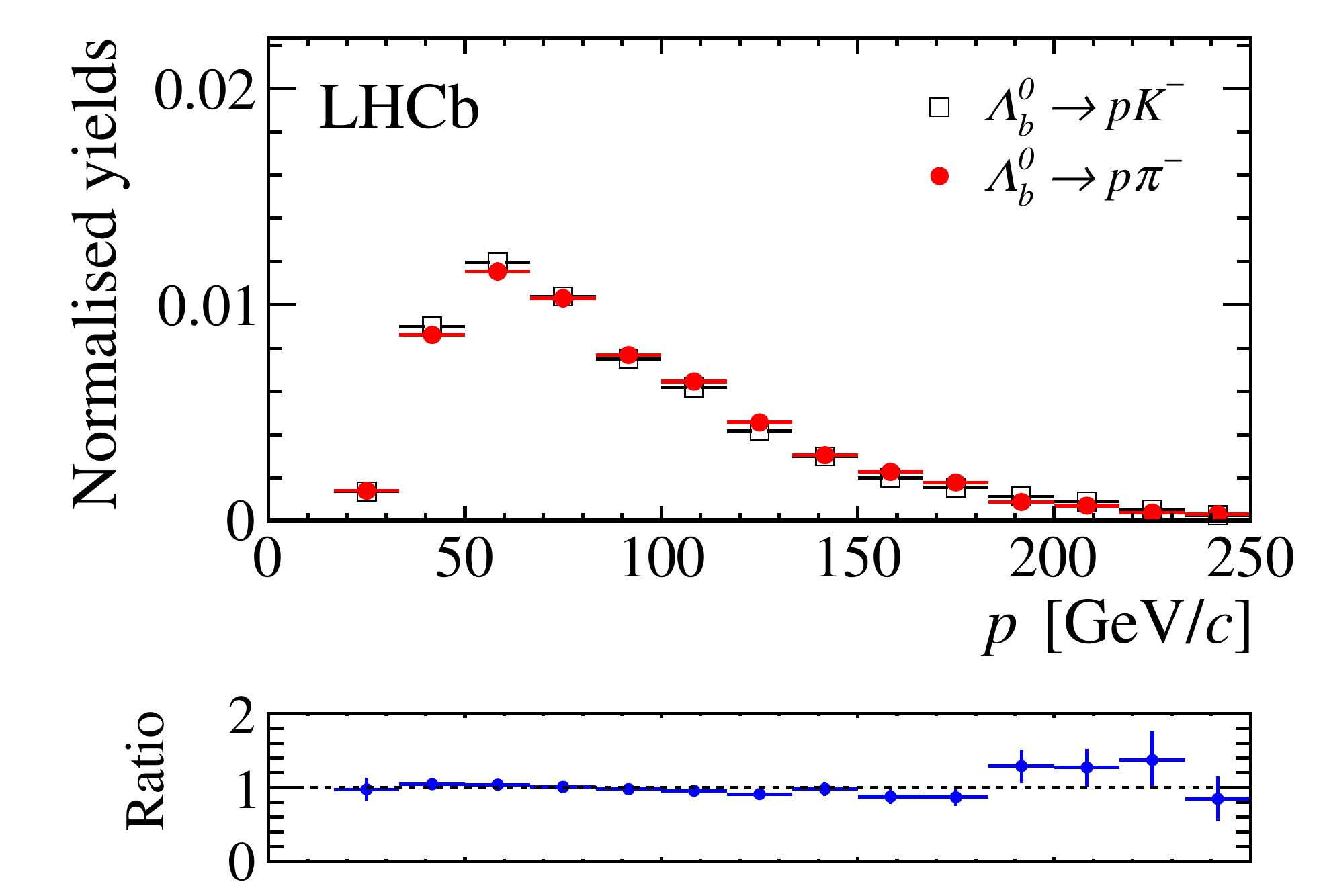}\\
\includegraphics[width=0.49\textwidth]{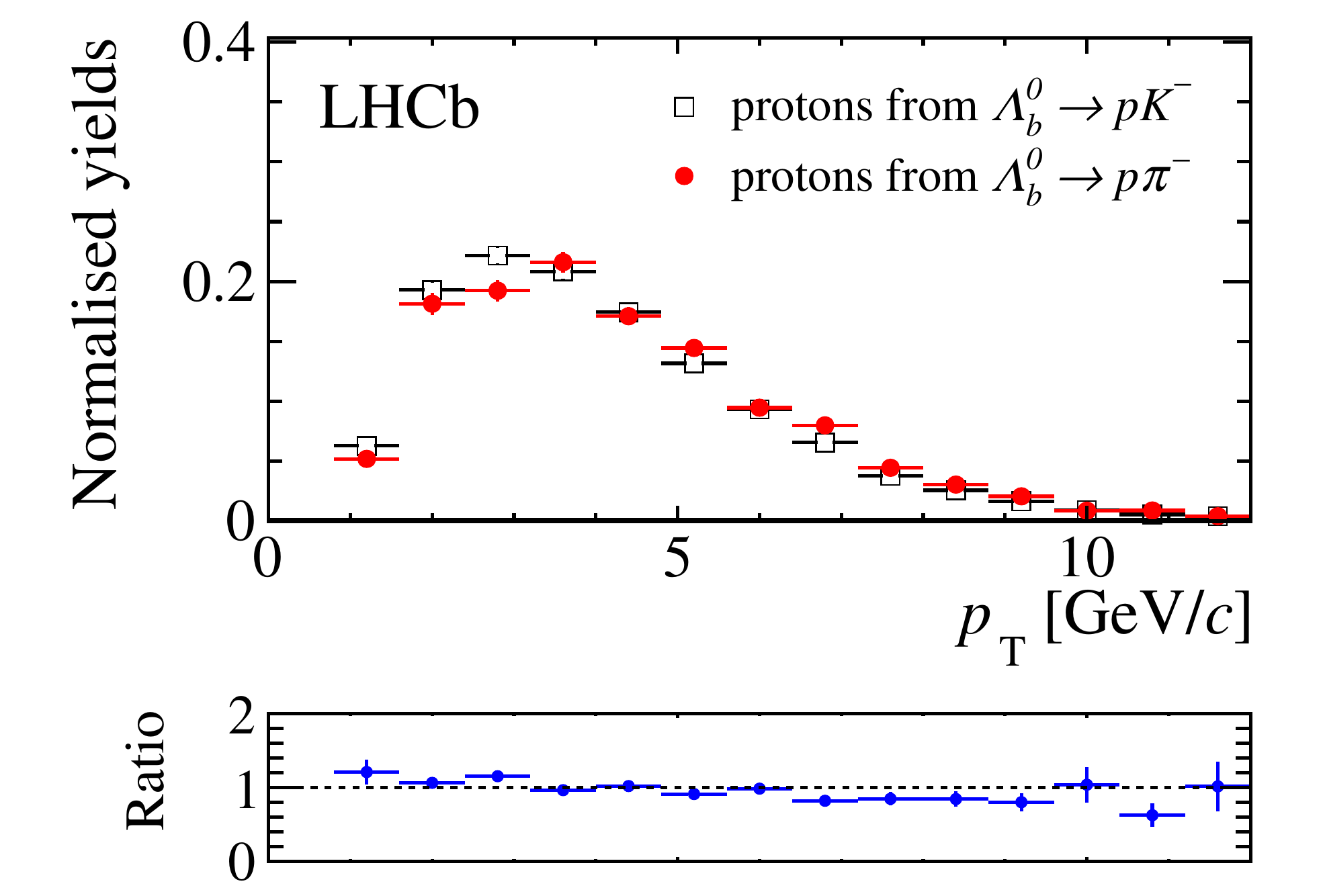}
\includegraphics[width=0.49\textwidth]{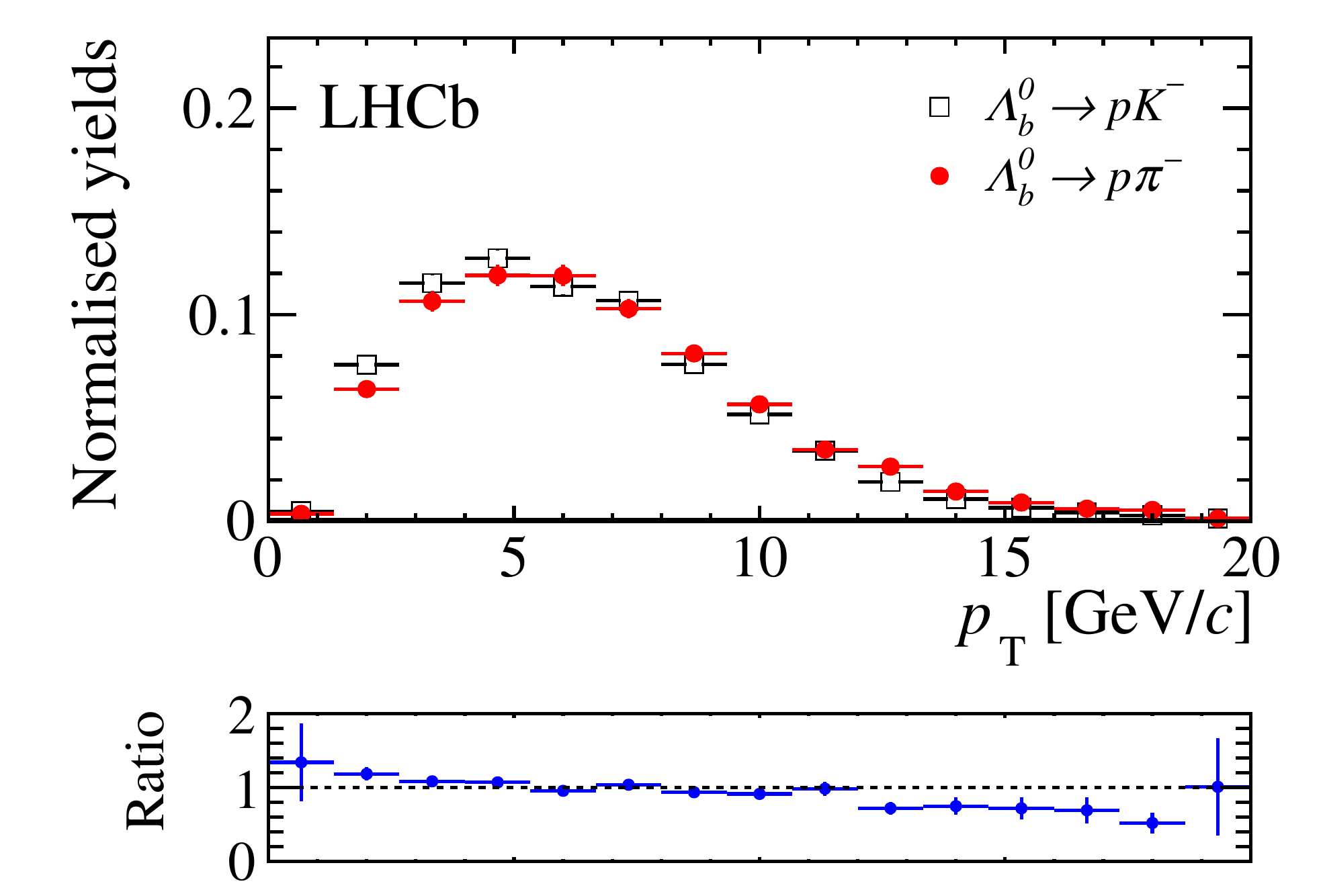}\\
\includegraphics[width=0.49\textwidth]{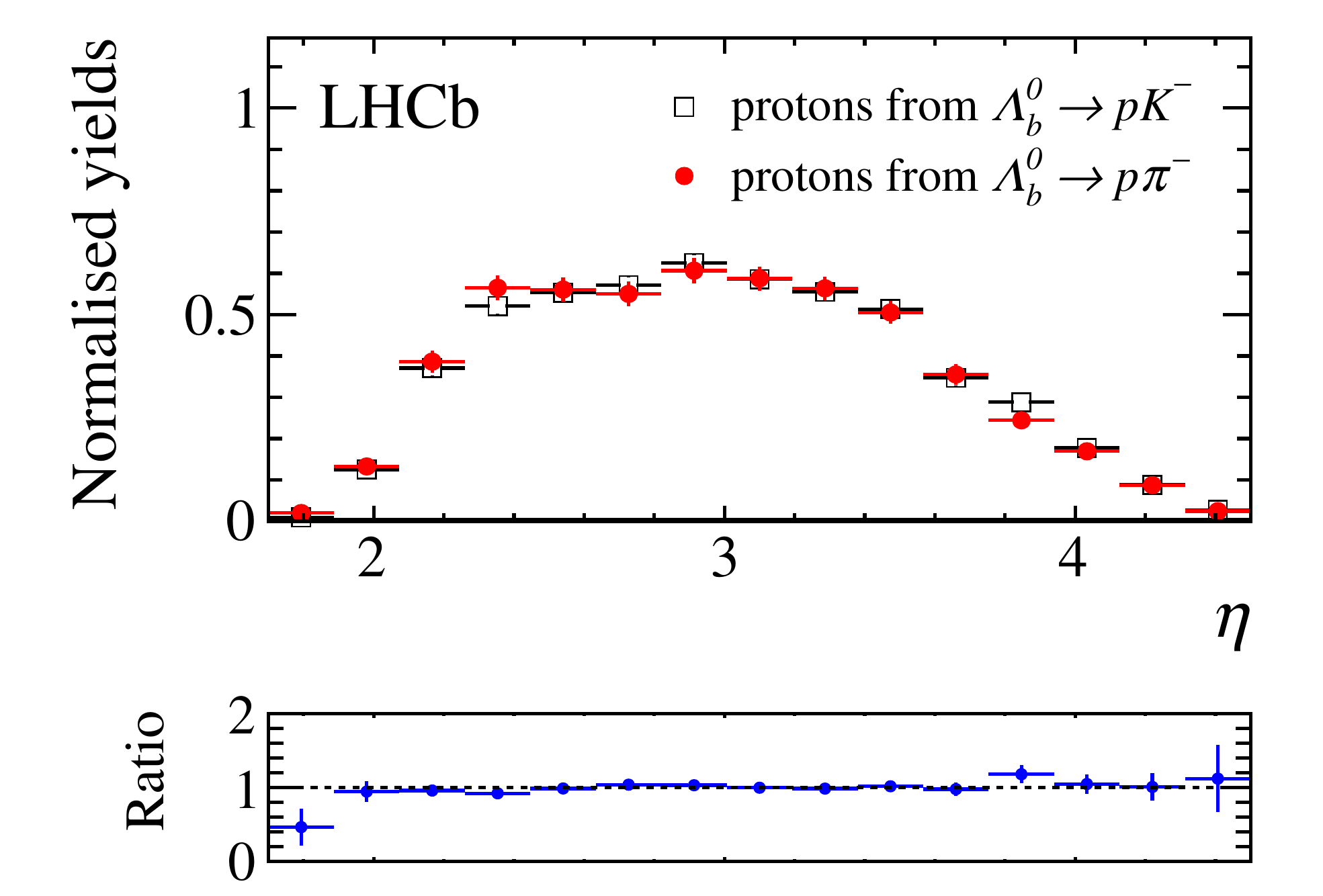}
\includegraphics[width=0.49\textwidth]{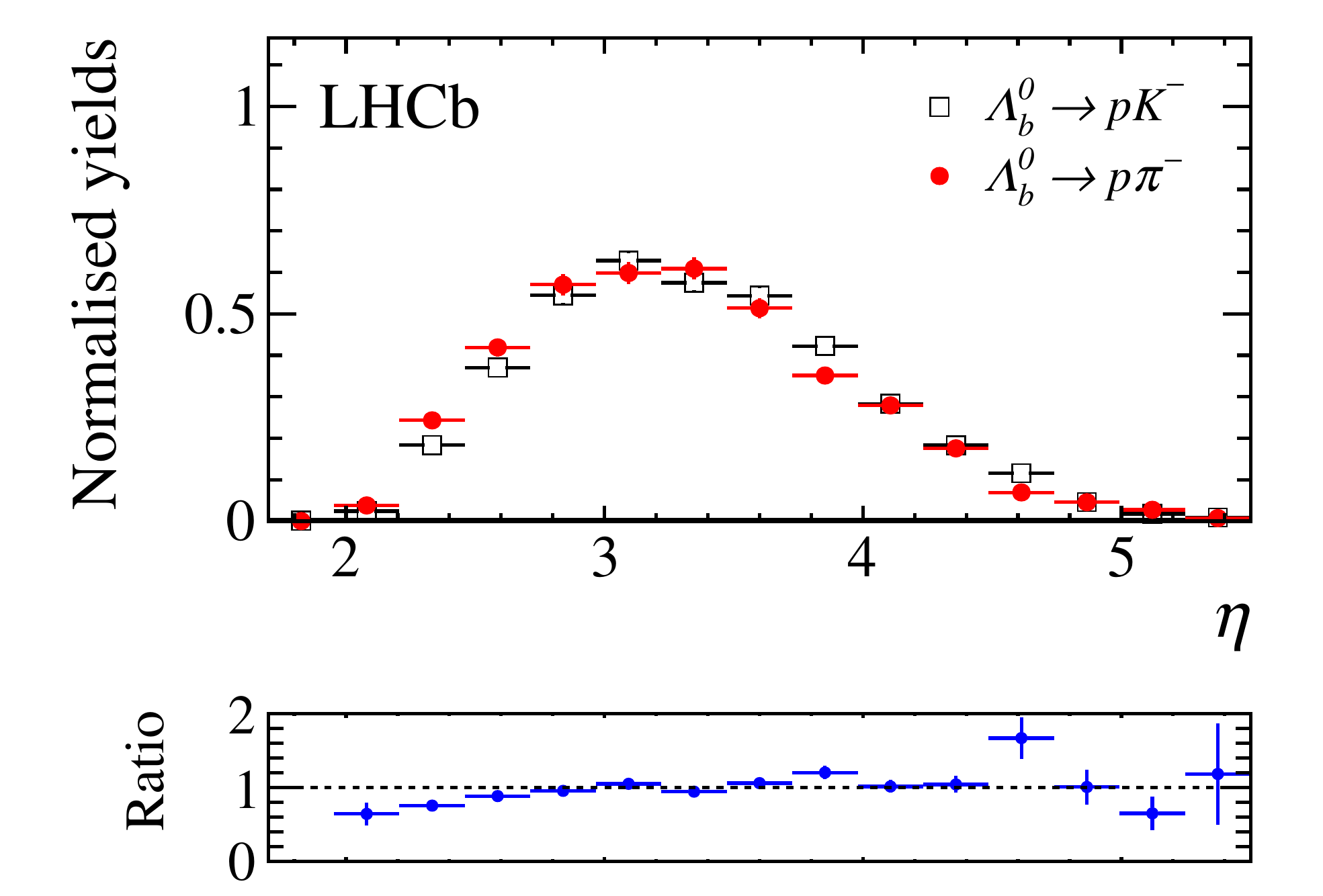}
\caption{Distributions of (top) momentum, (middle) transverse momentum and (bottom) pseudorapidity for (left) protons from \Lb decays and (right) \Lb baryons. The distributions are background-subtracted and normalised to unit area. Below each plot the ratio between the two distributions corresponding to \LbTopK and \LbToppi decays is also shown.}
\label{fig:KinDistrComparisonproton}
\end{figure}

The kaon detection asymmetry is determined as a function of the kaon momentum, following the approach developed in Ref.~\cite{LHCb-PAPER-2014-013} and subtracting $\ADKz=(0.054\pm0.014)\%$~\cite{LHCb-PAPER-2014-013} and the pion detection asymmetry. The momentum-dependent values are then weighted with the background-subtracted~\cite{Pivk:2004ty} momentum distribution of kaons from \LbTopK decays to obtain $\ADKm = (-0.76 \pm 0.23)\%$, where the dominant uncertainty is due to the finite size of the samples used. The pion detection asymmetry is obtained in an analogous way, adopting the approach of Ref.~\cite{LHCb-PAPER-2012-009}, and is determined to be $\ADpim = (0.13 \pm 0.11)\%$.
A different approach is followed for the proton detection asymmetry, since no measurement of this quantity is available to date. Simulated events are used to obtain the reconstruction efficiency defined as the number of reconstructed over generated decays, in bins of proton momentum. Then, according to Eq.~\eqref{eq:DetectionAsymmetries}, an asymmetry is defined and weights are computed from the background-subtracted~\cite{Pivk:2004ty} proton-momentum distributions of \LbTopK and \LbToppi decays. The proton detection asymmetries for both decays are found to be equal, consistent with the fact that the kinematics of the protons for the two decays do not exhibit significant differences, as shown in Fig.~\ref{fig:KinDistrComparisonproton}. The common value is $A_\mathrm{D}^\proton = (1.30 \pm 0.03 \pm 0.16 \pm 0.65)\%$, where the first uncertainty is due to the finite amount of simulated events and the second is associated to the knowledge of the material budget of the \lhcb detector. The third uncertainty is due to the assumptions made on the proton and antiproton cross-sections used in the computation.

The PID asymmetries are calculated using calibration samples with the aid of simulation to account for the limited phase-space coverage of the protons from $\Lz \to \proton \pim$ and $\Lc \to \proton \Km \pip$ decays. The dominant uncertainty comes from different PID performances in data and simulation in the phase-space region where simulated events are used. This discrepancy has been studied using \BdToKpi decays, for which the phase-space coverage of calibration data is larger. The values of the PID asymmetries are found to be $\APIDpK = (-0.30 \pm 0.74)\%$ and $\APIDppi = (-0.18 \pm 0.73)\%$.

The integrated \Lb production asymmetries are calculated convolving the background-subtracted~\cite{Pivk:2004ty} two-dimensional transverse-momentum and rapidity distributions of \mbox{\LbTopK} and \LbToppi candidates with the production asymmetries measured as a function of the same variables reported in Ref.~\cite{LHCb-PAPER-2016-062}. Since \Lb baryons selected in the $\proton\Km$ or $\proton\pim$ final states have very similar kinematics, as shown in Fig.~\ref{fig:KinDistrComparisonproton}, the value $\APLb = (2.7 \pm 1.4)\%$, averaged for 7 and 8\tev data, is obtained for the production asymmetry of both decays.

Asymmetries related to different trigger efficiencies for the charge-conjugated final states may arise. The efficiency for a charged hadron to be responsible for the affirmative decision of the hardware trigger is determined as a function of transverse momentum, separately for positively and negatively charged particles, using a sample of $\Lb \to \Lc(\to \proton \Km \pip) \pim$ decays. These efficiencies are used to determine the charge asymmetry introduced by the hardware trigger for the signal candidates that fire it. The charge asymmetry introduced by the hardware trigger for candidates that are retained independently of whether or not they are responsible for an affirmative hardware-trigger decision is determined studying a sample of \BdToKpi decays~\cite{LHCb-DP-2012-004}. The asymmetry of the software trigger is also studied using \BdToKpi decays, determining the charge asymmetry of the fraction of \BdToKpi decays for which both final-state hadrons fire the software trigger with respect to those for which only one hadron fires. The total trigger asymmetries are measured to be $\ATpK = (0.18\pm0.53)\%$ and $\ATppi = (-0.08\pm0.55)\%$. The uncertainties are mainly due to the limited size of the samples used in their determination.

Several sources of systematic uncertainties associated with the fit model are investigated. The alternative models used to determine systematic uncertainties associated with the choices of the invariant-mass shapes consist in turn of: adding a Gaussian function to the invariant-mass resolution model used for signals and cross-feed backgrounds to account for long tails due to candidates with a poor determination of the final-state particles momenta; changing the value of the parameter governing the final-state photon radiation effects according to its uncertainty; substituting the exponential function used to model the combinatorial background with a linear function and removing the partially reconstructed background component by rejecting candidates with $m_{\proton\Km}$~$(m_{\proton\pim})$ lower than $5.5$\gevcc.
When testing alternative models, 250 pseudoexperiments are generated according to the baseline fit model and using as input the central values of the baseline results. Fits are performed to each of the generated samples using the baseline model and then the alternative models. The mean and the root mean square of the distribution of the difference between the raw asymmetries determined by the two sets of fits are added in quadrature and the resulting value is taken as a systematic uncertainty.

 A different approach is adopted to assess systematic uncertainties related to the knowledge of the PID efficiencies. Samples are generated using the baseline fit model and results. The baseline fit model is then fitted 250 times to the generated samples, varying the PID efficiencies according to their uncertainties, which are mainly driven by the choice of the binning scheme used to divide the phase-space. These uncertainties are assessed by changing the baseline binning scheme with alternative schemes and computing again the efficiencies. The largest root mean square of the raw asymmetry distributions is taken as a systematic uncertainty.

 The systematic uncertainties due to the fit model choice, PID efficiencies determination and instrumental asymmetries measurement, along with the total uncertainty obtained  as the quadratic sum of the individual contributions, are reported in Table~\ref{tab:systematic}.
\begin{table}[!tbp]
  \caption{Systematic uncertainties on \ACPpK and \ACPppi.}
    \vspace{-0.5cm}
 \begin{center}
 \begin{tabular}{l|c|c}
        Systematic uncertainty & \ACPpK [\%] & \ACPppi [\%]\\
      \hline
        Kaon or pion detection asymmetry&  $ 0.23 $ & $ 0.11 $ \\
        Proton detection asymmetry &       $ 0.67 $ & $ 0.67 $ \\
        PID asymmetry &                    $ 0.74 $ & $ 0.73 $ \\
        \Lb production asymmetry &         $ 1.40 $ & $ 1.40 $ \\
        Trigger asymmetry &                $ 0.53 $ & $ 0.55 $\\
        Signal model  &                    $ 0.02 $ & $ 0.02 $\\
        Background model &                 $ 0.23 $ & $ 0.47 $\\
        PID efficiencies &                 $ 0.57 $ & $ 0.74 $\\
      \hline
        Total &                            $ 1.91 $ & $ 2.00 $\\
    \end{tabular}
    \end{center}
  \label{tab:systematic}
\end{table}

\section{Results and conclusions}
\label{sec:ResultsAndConclusions}
Using in Eqs.~\eqref{eq:ACPpK} and~\eqref{eq:ACPppi} the values of the raw asymmetries reported in Sec.~\ref{sec:InvariantMassFit} and those of the instrumental and production asymmetries reported in Sec.~\ref{sec:InstrumentalAsymmetriesAndSystematicUncertainties}, the following \CP asymmetries are obtained
\begin{eqnarray}
\ACPpK &=& -0.020 \pm 0.013 \pm 0.019, \nonumber \\
\ACPppi &=& -0.035 \pm 0.017 \pm 0.020, \nonumber
\end{eqnarray}
where the first uncertainties are statistical and the second systematic. The correlation between \ACPpK and \ACPppi is found to be 0.5. No evidence for \CP violation is observed.

A quantity that is independent from the proton detection and \Lb production asymmetries is obtained by taking the difference
\begin{eqnarray}
\label{eq:DACP}
\dacp \equiv \ACPpK - \ACPppi &=& \ArawpK - \ADKm - \APIDpK - \ATpK \nonumber \\
                              &-& \Arawppi + \ADpim + \APIDppi + \ATppi.
\end{eqnarray}

The statistical and systematic correlations between the raw asymmetries, the PID asymmetries and the detection asymmetries are taken into account when propagating the uncertainty to \dacp, obtaining
\begin{equation}
\dacp = 0.014 \pm 0.022 \pm 0.010, \nonumber
\end{equation}
where the first uncertainty is statistical and the second systematic. These results represent the world's best measurements to date, with much improved precision with respect to previous \cdf determinations~\cite{Aaltonen:2014vra}.

\section*{Acknowledgements}
%
%
\noindent We express our gratitude to our colleagues in the CERN
accelerator departments for the excellent performance of the LHC. We
thank the technical and administrative staff at the LHCb
institutes.
We acknowledge support from CERN and from the national agencies:
CAPES, CNPq, FAPERJ and FINEP (Brazil);
MOST and NSFC (China);
CNRS/IN2P3 (France);
BMBF, DFG and MPG (Germany);
INFN (Italy);
NWO (Netherlands);
MNiSW and NCN (Poland);
MEN/IFA (Romania);
MinES and FASO (Russia);
MinECo (Spain);
SNSF and SER (Switzerland);
NASU (Ukraine);
STFC (United Kingdom);
NSF (USA).
We acknowledge the computing resources that are provided by CERN, IN2P3
(France), KIT and DESY (Germany), INFN (Italy), SURF (Netherlands),
PIC (Spain), GridPP (United Kingdom), RRCKI and Yandex
LLC (Russia), CSCS (Switzerland), IFIN-HH (Romania), CBPF (Brazil),
PL-GRID (Poland) and OSC (USA).
We are indebted to the communities behind the multiple open-source
software packages on which we depend.
Individual groups or members have received support from
AvH Foundation (Germany);
EPLANET, Marie Sk\l{}odowska-Curie Actions and ERC (European Union);
ANR, Labex P2IO and OCEVU, and R\'{e}gion Auvergne-Rh\^{o}ne-Alpes (France);
Key Research Program of Frontier Sciences of CAS, CAS PIFI, and the Thousand Talents Program (China);
RFBR, RSF and Yandex LLC (Russia);
GVA, XuntaGal and GENCAT (Spain);
the Royal Society
and the Leverhulme Trust (United Kingdom);
Laboratory Directed Research and Development program of LANL (USA).



\addcontentsline{toc}{section}{References}
\setboolean{inbibliography}{true}
\bibliographystyle{LHCb}
\bibliography{main,LHCb-PAPER,LHCb-CONF,LHCb-DP,LHCb-TDR}

\newpage


\newpage
\centerline{\large\bf LHCb collaboration}
\begin{flushleft}
\small
R.~Aaij$^{27}$,
B.~Adeva$^{41}$,
M.~Adinolfi$^{48}$,
C.A.~Aidala$^{73}$,
Z.~Ajaltouni$^{5}$,
S.~Akar$^{59}$,
P.~Albicocco$^{18}$,
J.~Albrecht$^{10}$,
F.~Alessio$^{42}$,
M.~Alexander$^{53}$
% ...

{\footnotesize \it
% ...
$ ^{5}$Clermont Universit{\'e}, Universit{\'e} Blaise Pascal, CNRS/IN2P3, LPC, Clermont-Ferrand, France\\
% ...
$ ^{10}$Fakult{\"a}t Physik, Technische Universit{\"a}t Dortmund, Dortmund, Germany\\
% ...
$ ^{18}$INFN Laboratori Nazionali di Frascati, Frascati, Italy\\
% ...
$ ^{27}$Nikhef National Institute for Subatomic Physics, Amsterdam, Netherlands\\
% ...
$ ^{41}$Instituto Galego de F{\'\i}sica de Altas Enerx{\'\i}as (IGFAE), Universidade de Santiago de Compostela, Santiago de Compostela, Spain\\
$ ^{42}$European Organization for Nuclear Research (CERN), Geneva, Switzerland\\
% ...
$ ^{48}$H.H. Wills Physics Laboratory, University of Bristol, Bristol, United Kingdom\\
% ...
$ ^{53}$School of Physics and Astronomy, University of Glasgow, Glasgow, United Kingdom\\
% ...
$ ^{59}$University of Cincinnati, Cincinnati, OH, United States\\
% ...
$ ^{61}$Syracuse University, Syracuse, NY, United States\\
% ...
$ ^{73}$University of Michigan, Ann Arbor, United States, associated to $^{61}$\\
% ...
}
\end{flushleft}

\end{document}